INTRODUCTION TO THE REPORT "INTERLANGUAGES AND SYNCHRONIC MODELS OF COMPUTATION."

A.V. Berka.
Isynchronise Ltd.
email: alex.berka@isynchronise.com

Abstract.

It is suggested that natural language has major structural defects, and is inappropriate as a template for formal and programming languages, and as a basis for deriving models of computation. A novel language system has given rise to promising alternatives to standard formal and processor network models of computation. A textual structure called an *interstring* is proposed. When linked with an abstract machine environment, an interstring shares sub-expressions, transfers data, and spatially allocates resources for the parallel evaluation of dataflow. Formal models called the α-*Ram* family are introduced, designed to support interstring programming languages (*interlanguages*). Distinct from dataflow, graph rewriting, and FPGA models, α-Ram instructions are bit level and execute in situ. They support sequential and parallel languages without the space/time overheads associated with the Turing Machine and λ-calculus, enabling massive programs to be simulated. The devices of one α-Ram model, called the *Synchronic A-Ram*, are fully connected and simpler than FPGA LUT's. A compiler for an interlanguage called *Space*, has been developed for the Synchronic A-Ram. Space is MIMD. strictly typed, and deterministic. Barring memory allocation and compilation, modules are referentially transparent. At a high level of abstraction, modules exhibit a state transition system, aiding verification. Data structures and parallel iteration are straightforward to implement, and allocations of sub-processes and data transfers to resources are implicit. Space points towards highly connected architectures called *Synchronic Engines*, that scale in a Globally Asynchronous Locally Synchronous manner. Synchronic Engines are more general purpose than systolic arrays and GPUs, and bypass programmability and resource conflict issues associated with multicores. If massive intra chip, wave-based interconnectivity with nanosecond reconfigurability becomes available, Synchronic Engines will be in favourable position to contend for the TOP500 parallel machines.

I. INTRODUCTION.

Consider the hypothesis that trees and graphs have not in themselves alone, revealed an optimal linguistic environment in which to represent formal structures that possess shared parts, and require some form of computation or transformation, such as dataflow. The current work may be summarised as an attempt to identify such an environment, and then to use it as a foundation for a novel computational paradigm, incorporating low level and intermediate formal models, up to and including massively parallel programming models and machine architectures. Described in the main report, the implementation of a viable, general purpose parallel programming environment on top of a simple, highly connected formal model of computation, without excessive space or time overheads, provides a foundational framework for reconfigurable synchronous digital circuits, and coarse grained arrays of ALUs (CGAs). In so doing, an alternative to the systolic approach to programming and controlling CGAs is attained, which delivers a novel paradigm of general purpose, high performance computation.

The report questions two outlooks associated with the multi-processor paradigm of parallel computing. Firstly, that the Von Neumann sequential thread and architectural model, are suitable building blocks respectively, for a general purpose parallel programming model, and a parallel computing architecture. Secondly, that the absence of faster than light communication, suggests that asynchrony and non-determinism are fundamental to parallel programming frameworks. Without originally intending to do so, the consideration of linguistic issues has led to an espousal for synchronous and deterministic approaches to parallel programming, and highly connected aggregates of ALUs as parallel architectures.

Henceforth all chapter and section references relate to the main report, which can be downloaded via links on www.isynchronise.com. In chapter 8, a set of mostly synchronous architectural models with low area complexity high speed interconnects called *Synchronic Engines* are outlined, possessing spatially distributed, yet deterministic program control. Synchronic Engines are embryonic efforts at deriving architectures from a formal model of computation called the *Synchronic A-Ram* defined in chapter 3, inspired by interstrings and the interlanguage environment presented in chapter 2.

An interstring is a set-theoretical construct, designed for describing many-to-many relationships, dataflows, and simultaneous processes. It may be represented as a string of strings of symbol strings, where the innermost strings are short and have a maximum length. Interstring syntax is confined to a strictly limited range of tree forms, where only the rightmost, and the set of rightmost but one branches are indefinitely extendable. In conjunction with an abstract machine environment that does not reference semantics, an interstring can efficiently express sharing of subexpressions in a dataflow, data transfers, spatial allocation of machine resources, and program control for the parallel processing of complex programs. Languages based on interstrings are called *interlanguages*[1]. Although not incorporated in the current implementation, an interlanguage compiler may duplicate the implicit parallelism of Dataflow Models (see 2.3.3), where arithmetic operations from differing layers in a dataflow are triggered simultaneously, if outputs from operations in earlier layers become available soon enough.

In contrast with dataflow and visual programming formalisms, interlanguages are purely textual, making them directly amenable for digital representation and manipulation. The report explains how interlanguages, and more generally interlanguages based on more deeply nested string structures, where some inner strings are restricted to having a maximum length, are also useful for representing data structures intended to be processed in parallel.

---

[1] Interlanguages referred to here, have no relation to Selinker's linguistics notion of second natural language acquisition.

The Synchronic A-Ram is a globally clocked, fine grained, simultaneous read, exclusive write machine. It incorporates a large array of registers, wherein the transmission of information between any two registers or bits occurs in constant time. Although problematic from a physical standpoint, it will be argued that this assumption facilitates a conceptual advance in organising parallel processing, and can be worked around in the derivation of feasible architectures by various means, including the use of emerging wave based interconnect technologies, and permitting differing propagation delays across variable distances within a synchronous domain. Less optimal, purely wire based platforms, and globally asynchronous, locally synchronous (GALS) strategies may also be considered.

In a succession of Synchronic A-Ram machine cycles, an evolving subset of registers are active. Subject to some restrictions, any register is capable of either holding data, or of executing one of four primitive instructions in a cycle: the first two involve writing either '0' or '1' to any bit in the register array, identified by instruction operands, the third instructs the register to inspect any bit in the register array, and select either the next or next but one register for activation in the following machine cycle, and the fourth is a jump which can activate the instruction in any register in the following machine cycle, and also those in subsequent registers specified by an offset operand. Whilst the model's normal operation is relatively simple to explain, it's formal definition incorporates error conditions, and is somewhat more involved than that of a Turing Machine.

In common with assembly languages, schematic representations used for VLSI design and programming FPGAs, the hardware description languages VHDL and Verilog, and configuration software for systolic dataflow [1] [2] in coarse grained reconfigurable architectures, interlanguages may be characterised as *spatially oriented*. A programming language is spatially oriented if (i) there is some associated machine environment abstract or otherwise, and (ii) a program instruction or module, is linked in some way before runtime with that part of the machine environment, in which it will be executed in.

Vahid [3] and Hartenstein [4] stress the need for educators to consider spatially oriented languages, as important as conventional, non-spatial software languages in computer science curricula, because they are fundamental for expressing digital circuits, dataflows and parallel processes generally. The attitude that software and hardware may be studied in isolation from each other, is profoundly misguided. This report contains an account of how a high level, spatial language can easily deal with communication, scheduling, and resource allocation issues in parallel computing, by resolving them explicitly in an incremental manner, module by module, whilst ascending the ladder of abstraction. In what is in my view the abscence of viable alternatives, it can be conjectured that parallel languages *have* to be spatial. In 1.2, it is discussed how an non-spatial language and compiler system that attempts to deal with allocation and contention implicitly, is subject to a particular kind of state explosion, resulting from transforming a collection of high level non-spatial processes, into the lowest level, machine-bound actions. Lee in [13] argues non-deterministic multi-threading introduces another kind of state explosion, making the establishment of program equivalence between threads intractable.

*Space* is a programming interlanguage for the Synchronic A-Ram, and may describe algorithms at any level of abstraction, with the temporary exceptions of virtual functions and abstract data types. Moreover, it is possible to incorporate parallel iteration and typed data structures, without adding the overheads and deadlocks to programs, that are associated with conventional dataflow or graph based programming environments (see 2.3.3 and 2.3.4.) An interlanguage compiler produces code that at runtime, is capable of generating massive operational parallelism at every level of abstraction.

Providing a simple programming methodology is adhered to, Space's runtime environment, perhaps surprisingly, does not need to consider resource contention, deadlocks, and Synchronic A-Ram machine errors, because these issues have been implicitly dealt with at compile time. Race and time hazards are resolved by local synchronisation mechanisms. These features are scalable, and conceptually represent significant advantages over multi-threading on processor networks.

II INTERCONNECT AND SYNCHRONISATION TECHNOLOGIES, AND RELATED WORK IN RECONFIGURABLE COMPUTING.

Reference is made to David Miller's work in 1.2.2, on using light as a means of synchronising room sized systems to nanosecond/picosecond intervals, of relevance to the construction of very large, globally clocked computers. In 8.3, the prospects of implementing a highly interconnected massive array of small computational elements, using either an optically or spintronically based network architecture are discussed. In 8.4, it is also explained how global synchrony can be relaxed in Synchronic Engines, to allow greater scalability. Massively parallel programs would still be conceived as globally clocked processes, aiding programmability, but would to a large extent run asynchronously.

The apparent lack of wave-based intra-chip connections allowing reconfigurable connectivity on the order of nanoseconds, indicates that more efficient Synchronic Engines may not be fully realisable in the short to mid term. In 8.2.1, a photonic connection system is described, in which microsecond switching between large numbers of nodes without chip area explosion, seems within reach. In 8.2.2, a spin-wave technology is outlined, that may enable nanosecond data exchange times for nano architectures incorporating millions of devices. A comparison between interlanguage programming on currently buildable Synchronic Engines, and multi-threading on multi-processor networks on standard industry benchmarks, will become available further down the research path.

The consideration of using silicon alone to realise less efficient machines, revealed a close relationship between the current approach and the field of reconfigurable computing, which was only fully appreciated in the final stages of writing this report. The action of a Synchronic A-Ram register is more primitive than a logic gate or FPGA look up table, and the register array's bits are in a sense, fully connected. It will be argued in a future paper, that if propagation delay were introduced into the definition (see 3.5.2), the model is fundamental to physical reconfigurable computing. Synchronic A-Rams are finer grained and more connected, and may therefore simulate FPGAs and CGAs without the inefficiencies that conventional reconfigurable models would have simulating each other.

Further, spatial computation based on systolic processing, on grids of coarse grained functional units, that might be termed *systolic spatialism,* lacks an abstract model, beyond the coarse grained, systolic grid itself. The approach suffers from being domain restricted; the developer is obliged to cast every program as a Digital Signal Processing-like collection of pipes or streams [5]. Systolic spatialism is however, well matched to silicon's restricted, planar connectivity.2 It is an effective approach for maximising utilization and performance in wire-based parallel architectures, for applications that can be cast as streams [1] [2] [7].

Interlanguages form the basis for developing a new class of more general purpose programming models for wire based FPGAs and Coarse Grained Arrays of ALUs. There is a concern that the interlanguage model might lead to lower efficiency of runtime resource utilisation compared with purely systolic approaches, unless compensatory mechanisms are introduced (see 8.4).

Alternative kinds of programming environments for FPGA and reconfigurable platforms require a significant amount of hardware expertise from the developer [6], do not port to new architectures [7], and do not adequately support general purpose parallelism [8]. Sequential language environments for reconfigurable platforms might offer the prospect of parallelizing the software base, but by their nature do not allow the expression of parallel algorithmics. Their compilers [9] [10] rely on reassembling dataflows from arithmetic operations and loop unrolling, for parallelization. They cannot transform inherently sequential algorithms, which might appear anywhere in the spectrum of abstraction, into efficient parallel programs. Languages that do offer extensions for multi-threading on reconfigurable fabrics [11] [12], inherit the limitations of multi-threading (see the next section).

The authors in [3] [4] stress the severe overheads arising out of instruction fetch in processor networks, that are bypassed in spatial computing, because instructions are executed in situ. In 1.2, the case against processor networks is further examined, in that they lack a good high level programming model and theoretical basis. The impact of their ubiquity in fields of application is discussed.

III A NEW APPROACH TO LANGUAGE AND COMPUTATION.

A more detailed overview of the report follows. The historical development of human language has not been optimal, for it's use as a template for formal and programming languages. Tree syntax is common to all natural languages, and has three structural defects. The first defect is called the Single Parent Restriction (SPR), and relates to the expression of many-to-many relationships. SPR is the linguistic counterpart to the defining characteristic of conventional tree syntax; every node or part of speech may only participate in at most one more complex part of speech. SPR limits a part of speech describing an object, from participating directly in the expression of more than one relationship on a syntactic level, unless some form of naming, normally involving a semantic notion of storage, or sub-expression repetition is used. Repetition results in a potentially exponential increase in size for dataflow representations with respect to dataflow depth, compared with graph forms [9]. SPR is also associated with trees with high structural variability in which any branch may be arbitrarily long, requiring complex parsing, and whose contents cannot easily be identified and accessed in parallel.

A second defect relates to natural languages inability to express parallelism directly, so that many basic sentence representations may be conveyed simultaneously, thereby potentially providing a cue for their meanings to be processed simultaneously.

The third defect relates to natural languages non-spatial character. They do not allow the expression of abstract spatial information at the level of syntax, relating to data transfers and allocation of jobs to resources on the semantic level, that is argued in the report to be helpful in avoiding resource contention and state explosion in general purpose parallel computing. The use of dynamic semantics alone to deal with the effects of the three defects, represent a partial solution which discards an opportunity to devise a better general purpose language structure.

The emergence of the non-spatial tree, as the de facto, standard language structure for syntax and semantics, has had serious consequences for our capacity to describe and reason about complex objects and situations. The inability to directly share subexpressions contributes to code bloat in commercial software, disconnected representations of environments, and a kind of linguistic schizophrenia. An unrestricted recursive application of rewriting rules for symbol strings is suboptimal linguistically, in that it is not conducive for describing simultaneous processes. Tree formalisms have deterred the introduction of an explicit notion of time and computation into mathematics.

The problematic nature of subexpression repetition in particular has been noticed before, and has given rise to graph/data flow models, such as Term Graph Rewriting, Petri Nets, Semantic Nets, and Dataflow models. But these approaches have not entered the mainstream. Although the basic structure used is that of a graph, they are described in conventional tree-based mathematics, involving the non-spatial transformation of expressions alone, and lack an explicit notion of a computational environment. They are implemented on networks of Turing Machines/processors, do not call for a fundamental rethink of formal models of computation, and rarely call for an alternative computer architecture to the processor network.

An alternative to conventional tree based syntax and semantics has been devised in the form of an a language environment called *interlanguage*. The environment consists of a language based on the notion of the interstring, and an abstract memory and functional unit array, capable of storing elements, and performing operations of some given algebra. Interlanguage allows the sharing of subexpressions to be explicitly represented, with linear cost with respect to the number of subexpressions. The tree form of an interstring is highly regular, requiring only a minimal syntactic analysis phase before semantic processing. Interstrings syntactically indicate which subexpressions may be semantically processed simultaneously, and allow resource allocation to be performed implicitly. Interstrings are also suitable for representing data structures with shared parts, and are intended to replace trees and graphs as standard programming structures.

The α-Ram family of machines are formal models of computation, which have been developed to be the target machines for the compilation of high level programs expressed in an interlanguage. Members of the α-Ram family with infinite memories are Turing Computable.

A member of the α-Ram family with finite memory, called the Synchronic A-Ram, may be viewed as a formal model underpinning the concept of an FPGA or reconfigurable machine. It supersedes finitistic versions of the Turing Machine and the λ-Calculus, the current standard models of Computer Science, in it's ability to efficiently support a high level parallel language. There is the prospect of a proper formalisation of parallel algorithmics, a new way of relating operational and denotational program semantics, and novel opportunities for parallel program verification. Massive instruction level parallelism can be supported, storage and processing resources are integrated at the lowest level, with a control mechanism similar to a safe Petri Net marking.

An interlanguage called *Space*, has been designed to run on the Synchronic A-Ram. Space is an easy to understand, fully general purpose parallel programming model, which shields the programmer from low level resource allocation and scheduling issues. Programs are textual rather than graphic, and iteration, data structures, and performance evaluation are supported. Space has a high level sequential state transition semantics, and solves the conceptual problem of how to orchestrate general purpose parallel computation, in a way that has not been achieved before.

The set-theoretical/logical definition of procedures for assembling constructions in mathematics, and the constructions themselves, are normally considered to reside in a universe of discourse, which is neutral and abstract from any computational implementation. A claim is made however, that conventional tree based formalisms in pure mathematics, harbour implicit notions of sequential, asynchronous and recursion oriented computation. Further, a universe of discourse incorporating an explicit parallel computational environment, is amenable to the adoption of parallel forms of reasoning, that bypass an implicitly sequential style in conventional mathematical discourse.

Synchronic Engines are physical architectural models derived from the Synchronic A-Ram and Space, and are composed of large arrays of fully, or extensively connected storage and processing elements. The models suggest optoelectronic, and spin-wave based hardware specifications. If interconnect issues can be overcome, there is a new avenue for developing programmable and efficient high performance architectures.

Without having yet provided detail, the class of Space-like interlanguages, and the associated formal and hardware platforms, which during execution preserve their parallelism and lack of resource contention, constitute a paradigm of parallel computation that will be termed *synchronic computation.*

IV SPACE.

Space is a programming interlanguage with a functionality comparable to C. Space programming has an applicative style, and bypasses the readability and efficiency issues associated with recursion based, functional style programming. In order to explore design issues arising from the interaction of interlanguage and machine resources, a Synchronic A-Ram simulator has been written, and a substantial software project has resulted in a programming environment called *Spatiale* (*Spatial E*nvironment) being developed. Spatiale is a non-GUI, unix console application written in C, and incorporates a compiler that transforms Space programs into Synchronic A-Ram machine code. The package and documentation are available via links on www.isynchronise.com.

Spatiale is intended to serve as a prototype for Synchronic Engine programming environments. Space would require little adaptation in order to program Synchronic Engines. It is an explicitly parallel, deterministic, strictly typed, largely referentially transparent language, that retains the notion of updateable states. Although the Space programmer is obliged to consider some scheduling and resource allocation issues, these are relatively transparent within the narrow, synchronous and deterministic context of a module, and he is shielded from issues pertaining to pre-defined modules. They have been resolved by earlier composition, leaving the compiler to implicitly perform these tasks at compilation time.

Space modules are not generally intended to retain states between activations. At the current stage of the compiler boot strapping process, a high degree of referential transparency can be attained. It cannot be unequivocably ascribed to Space, because the programmer is obliged to ensure a module resets it's internal values after execution. In addition, memory allocation and reconfigurable interconnect features are required to bridge the gap between a high level program environment and a low level machine. It is envisioned that later versions of Space will have built in support for low level mechanisms, that will guarantee referential transparency for new program modules.

In Space, as well as in the Synchronic A-Ram machine code, more than one simultaneous write to a storage area, and more than one simultaneous call to a processing resource, results in machine error. The error mechanisms do not appear to restrict the expression of deterministic parallel algorithms. Space modularisation and programming methodology, lead to the avoidance of race conditions and deadlocks, and enhanced software maintainability. The ability to modularise scheduling and resource allocation, and avoid resource contention, gives rise to programming models and architectures, which have decisive advantages over multi-threading for processor networks.

A deterministic Space program with simultaneous subprograms running in a synchronous (or virtually synchronous) environment, is much easier to understand than a non-deterministic, asynchronous network of Von Neumann processes. Space has the benefits of functional programming, such as modular construction and lack of side effects, despite having updateable states. In addition, there are not the stack related inefficiencies associated with recursive function based computing. In order to provide proof of concept for synchronic computation, a range of massively parallel high level programs have already been successfully run on the simulator with outputs as expected. This has, to the best of my knowledge, never been achieved before with a simulated formal model of computation.

An implementation of synchronic computation onto processor networks is conceivable. Parallel sub-processes could be broken down into coarse grained blocks, and then sequentialised to run individually on a core, in the hope that some parallel speedup is preserved. Unfortunately, this approach would likely lead to low utilization of the panoply of conventional processor resources, and poor performance overall. Fine grained processes would need to synchronise and communicate across non-adjacent cores, resulting in long waits for maintaining cache coherency, and for the interconnection network to transfer results, leaving ALUs idle for many machine cycles. In addition, interlanguages offer no obvious opportunities for exploiting the extensive hardware resources dedicated to supporting speculation, predication, and the elimination of race and time hazards for multiple, out of order instruction issue.

V ORGANISATION OF THE MAIN REPORT.

Chapter two justifies the introduction of the interlanguage environment, by comparing the ability of interstrings to represent dataflow and dataflow processing, with trees and graphs, and by providing a critique of historical attempts in Computer Science to deal with the structural defect of tree languages. Chapter three describes the α-Ram formal models of computation, inspired by interstrings. Chapter four defines a programming language called *Earth*, which is close to the Synchronic A-Ram machine code, and allows the definition of the most primitive program modules used in Space. Chapters 5 to 7 present the Space interlanguage itself. Space's type system and program declarations are laid out in chapter 5, and chapter 6 defines the basic interstring language structures, and presents some simple program examples. Chapter 7 covers programming constructs, enabling the description of massive parallelism, along with a range of program examples. In chapter 8, Synchronic Engines are presented. Chapter 9 discusses the relative merits of the standard models compared with α-Ram machines, and gives an outline of how efficiently simulable models offer new opportunities for unifying logic and mathematics with foundational computer science.